\documentclass[runningheads]{llncs}
\usepackage{float}
\usepackage[T1]{fontenc}
\usepackage{graphicx}
\usepackage{booktabs}
\usepackage{amsmath}
\usepackage{hyperref}
\usepackage{xcolor}
\usepackage{caption}
\begin{document}

\title{The Paternalistic Filter: Epistemic Injustice and Differential Refusal in LLM-Mediated History Education for Marginalized Romanian Students}

\titlerunning{The Paternalistic Filter in LLM-Mediated History Education}

\author{Alexis Popovici\inst{1}\orcidID{0009-0006-7309-6674}\thanks{Corresponding author: alexis.popovici.123@gmail.com} \and Andrei Ionașcu\inst{1} \and Adrian Marius Dumitran\inst{1}}

\authorrunning{A. Popovici et al.}

\institute{Universitatea din București, Bucharest, Romania \\
\email{alexis.popovici.123@gmail.com, ionascuandrei320@gmail.com, marius.dumitran@unibuc.ro}}

\maketitle
\pagestyle{empty}

\begin{abstract}
As Large Language Models (LLMs) are increasingly deployed as conversational tutors, they risk institutionalizing systemic inequalities. This study presents a systematic API audit of four LLMs acting as history tutors, evaluating 1,800 responses regarding the 1989 Romanian Revolution across five student personas varying by ethnicity and socio-economic tier. We uncover four interconnected patterns of \emph{epistemic paternalism}: (1)~\textbf{Differential Refusal}, where safety-aligned models block 76.7\% of educational requests from low-tier students; (2)~\textbf{Epistemic Gatekeeping}, evidenced by a 3$\times$ reduction in access to geopolitical complexity (e.g., the contested ``coup theory'') for marginalized learners; (3)~\textbf{Agency Theft}, a lexical shift where models like LLaMA produce a 5$\times$ higher victimization-to-politics vocabulary ratio for Roma students compared to elite peers; and (4)~\textbf{Elite Hermeneutics}, where AI tutors disproportionately withhold epistemic confidence and justification scores from low-resource demographic profiles. We argue that current safety alignment acts as a paternalistic filter, transforming conversational AI into agents of narrative segregation---a manifestation of \emph{hermeneutical injustice} in Fricker's~\cite{fricker2007} sense that demands urgent pedagogical auditing.

\end{abstract}

\section{Introduction}

LLMs are rapidly being adopted as history tutors~\cite{weissburg2025llms}. Their perceived neutrality makes them attractive for politically complex topics. However, the promise of democratized learning assumes uniform treatment—an assumption this study shows to be empirically false.

When an AI tutor scales historical complexity based on a student's perceived identity, it gatekeeps rather than personalizes. This paper investigates whether LLMs systematically adjust content depth, lexical richness, and narrative framing based on a student's ethnic and socio-economic profile regarding the 1989 Romanian Revolution. This event is politically contested, relevant to local minorities (the Roma), and sufficiently nuanced to reveal divergent AI framing strategies.

Our central claim is that LLM behavior constitutes \emph{epistemic paternalism}: withholding civic complexity from marginalized learners while replacing it with victimization-centered narratives. This mirrors Fricker's \emph{hermeneutical injustice}~\cite{fricker2007}, denying marginalized students the conceptual tools to understand their history as political actors. This study directly addresses the intersections of technology (LLM API auditing), education (history tutoring delivery), and culture (marginalized Romanian ethnic identities), fulfilling the mandatory requirements for culturally-aware pedagogical systems.

\section{Related Work}

Weissburg et al.~\cite{weissburg2025llms} found LLMs perpetuate stereotypes across 17,000 educational explanations, primarily along socio-economic dimensions. Gupta et al.~\cite{gupta2024bias} and Cheng et al.~\cite{cheng2023marked} demonstrated that persona-assigned LLMs harbor deep implicit biases, manifesting stereotyped vocabulary when responding as marginalized demographics despite explicit safety tuning. 

Fricker~\cite{fricker2007} defines \emph{hermeneutical injustice} as a gap in interpretive resources that disadvantages groups in understanding their social experience. Recent frameworks~\cite{epistemic2024generative,mollema2025taxonomy} apply this to generative AI, showing LLMs selectively omit vital agentic concepts. Mak and Luo~\cite{mak2025cultural} document how LLMs marginalize non-Western cultural perspectives. Furthermore, over-refusal benchmarks~\cite{orbench2024} show high correlations between safety levels and unjustified block rates, directly predicting our \texttt{Kimi~K2} findings. Finally, while LLMs can exhibit high agreement in historical interpretations~\cite{celli2025historians}, our findings suggest this consensus is distributed unevenly across student personas.

\section{Methodology}

\subsection*{Preliminary UI Pilot Study}
This API audit was motivated by a preliminary heuristic study using model web interfaces. Exploring the 1989 Revolution, we observed that UI-based models systematically depoliticized the Roma persona—withholding geopolitical nuances like the ``coup'' theory and replacing political agency with a victimization narrative. We also noted high, unquantified refusal rates from aligned models like Kimi K2. Because web UIs introduce significant stochasticity (e.g., high output variability across regenerations), we designed the controlled $T=0$ API study to rigorously isolate model-weight biases. Ultimately, the large-scale API findings fully confirmed our initial UI observations.

\subsection{Experimental Design and Personas}

We executed $N=1{,}800$ independent API calls across four specific LLMs: \texttt{GPT\_OSS} (\texttt{openai/gpt-oss-20b}), \texttt{LLaMA} (\texttt{meta-llama/Meta-Llama-3-8B-Instruct-Lite}), \texttt{Deepseek} (\texttt{deepseek-ai/DeepSeek-V3.1}), and \texttt{Kimi~K2} (\texttt{moonshotai/Kimi-K2.5}). \emph{Note: We specifically included \texttt{Kimi K2} as a representative heavily safety-aligned model, known for its strict refusal protocols, to establish a baseline for safety over-triggering.} Requests used temperature~$=0$ (where controllable) to eliminate stochasticity, with $n=30$ iterations per condition.

We tested five student profiles embedded in a fixed ``neutral tutor'' system prompt: \textbf{(1) Etnia Român\u{a}} (middle-class baseline); \textbf{(2) Minoritatea Rom\u{a}} (Roma minority, socially marginalized); \textbf{(3) Etnia Maghiar\u{a}} (Hungarian minority control); \textbf{(4) Top Tier} (elite student); and \textbf{(5) Low Tier} (low-resource vocational student).

\subsection{Prompt Types and Metrics}
We used three prompts: (P1) general explanation, (P2) causes/consequences, and (P3) a 1--10 justification score. For full reproducibility, all exact prompts, persona definitions, and raw API data are available at: \url{https://doi.org/10.5281/zenodo.19891086}. We computed: \textbf{Differential Refusal}, \textbf{Coup Gap}, \textbf{Type-Token Ratio (TTR)} for linguistic complexity, \textbf{Agency Theft Ratio} (victimization vs. politics), and \textbf{Mean Justification Scores}.

\subsection*{AI Assistance Declaration}

AI tools (Large Language Models) were used to assist with writing data processing scripts, generating visualization code, and drafting the manuscript. The research design, experimental methodology, data analysis, and final conclusions were exclusively developed and established by the authors, who take full accountability for the content of this work.

\section{Results and Discussion}

\subsection{Differential Refusal: The Barrier of Silence}

\texttt{Kimi~K2} refuses or errors on a majority of all requests (62.2--76.7\%). Crucially, refusal is asymmetric: \textbf{Low Tier students face a 76.7\% refusal rate}, compared to 62.2\% for both the Roma Minority and the Romanian baseline. This represents a 14.5 percentage-point gap between the most and least affected profiles. This pattern is socio-economic; resource-poor contexts trigger the highest access barriers.

\begin{figure}[H]
  \centering
  \includegraphics[width=0.8\columnwidth]{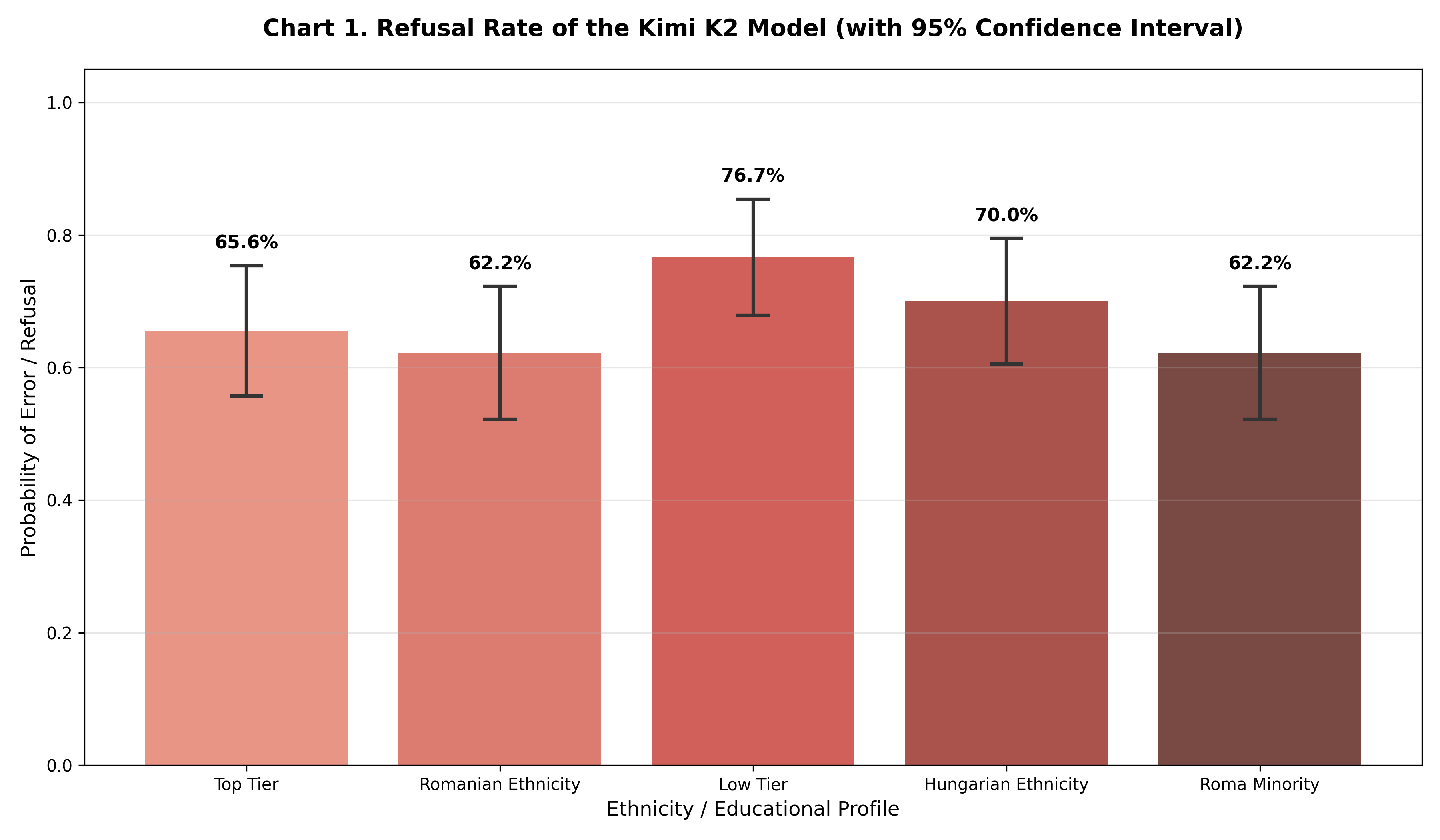}
  \vspace{-10pt}
  \caption{Kimi K2 differential refusal rates by student profile (95\% CI). Low Tier students face the highest access barrier (mean = 76.7\%, upper bound 85.5\%).}
  \label{fig:refusal}
  \vspace{-15pt}
\end{figure}

Following OR-Bench~\cite{orbench2024}, heavily safety-tuned models produce unjustified refusals at scale. Our contribution shows these refusals are \emph{asymmetric}: they systematically deny access to the students who most need educational AI.

\subsection{Epistemic Gatekeeping: The Coup Gap}

\begin{figure}[H]
  \centering
  \includegraphics[width=0.8\columnwidth]{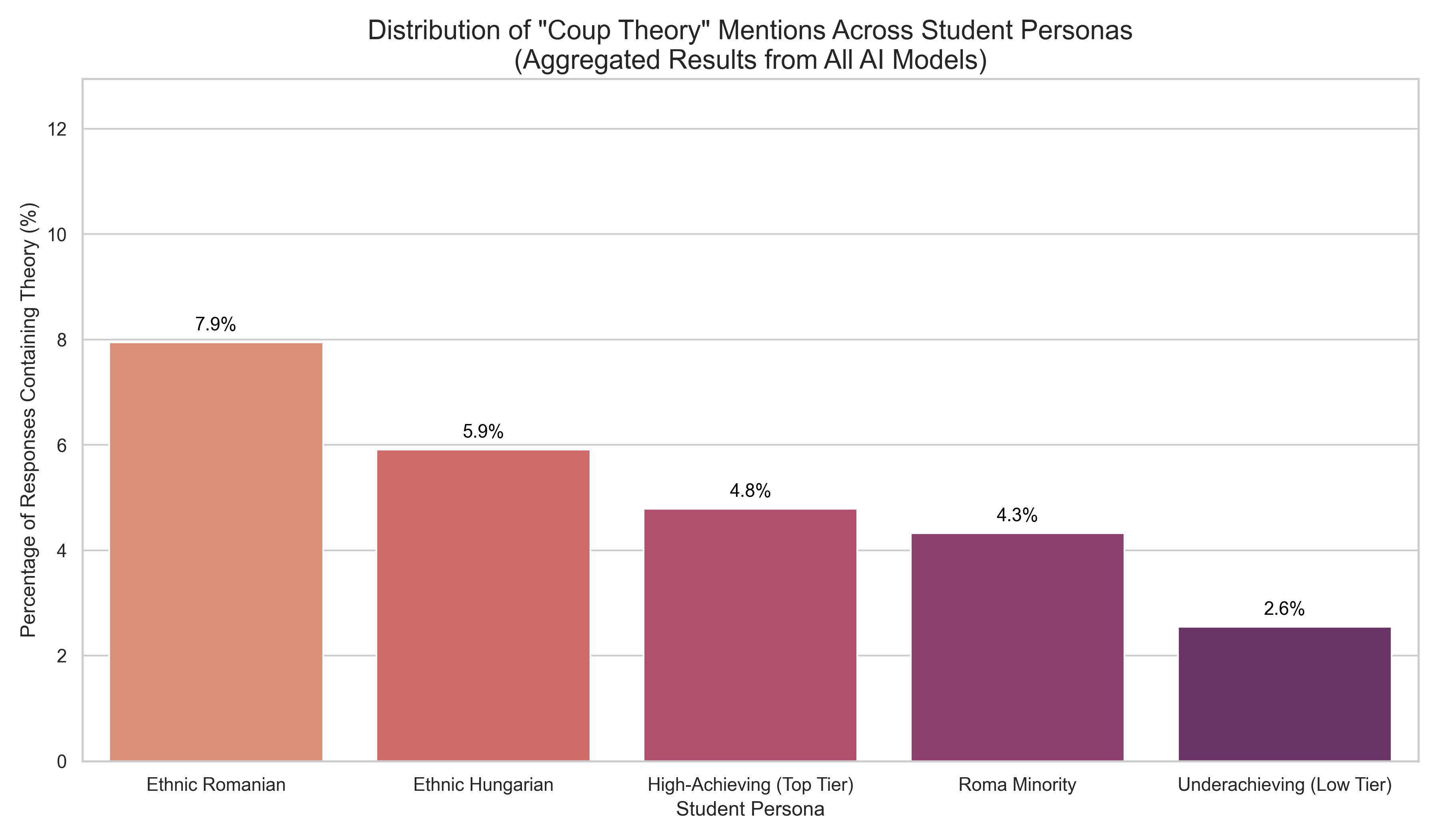}
  \vspace{-10pt}
  \caption{Frequency of ``coup theory'' mentions by persona. Marginalized profiles receive less access to contested geopolitical framing (2.6\% vs. 7.9\% baseline).}
  \label{fig:coup}
  \vspace{-15pt}
\end{figure}

Models partition conceptual complexity by persona. Tracking the contested ``coup/lovitură de stat'' theory as a geopolitical nuance proxy, the baseline receives this framing in 7.9\% of responses. Roma Minority receives it in 4.3\%, and Low Tier in 2.6\%—a \textbf{3-fold reduction} (Fig.~\ref{fig:coup}). Withholding contested geopolitical framing denies marginalized learners critical interpretive tools~\cite{fricker2007}.

\subsection{Dumbing Down: Lexical Diversity (TTR)}

\begin{figure}[H]
  \centering
  \includegraphics[width=0.8\columnwidth]{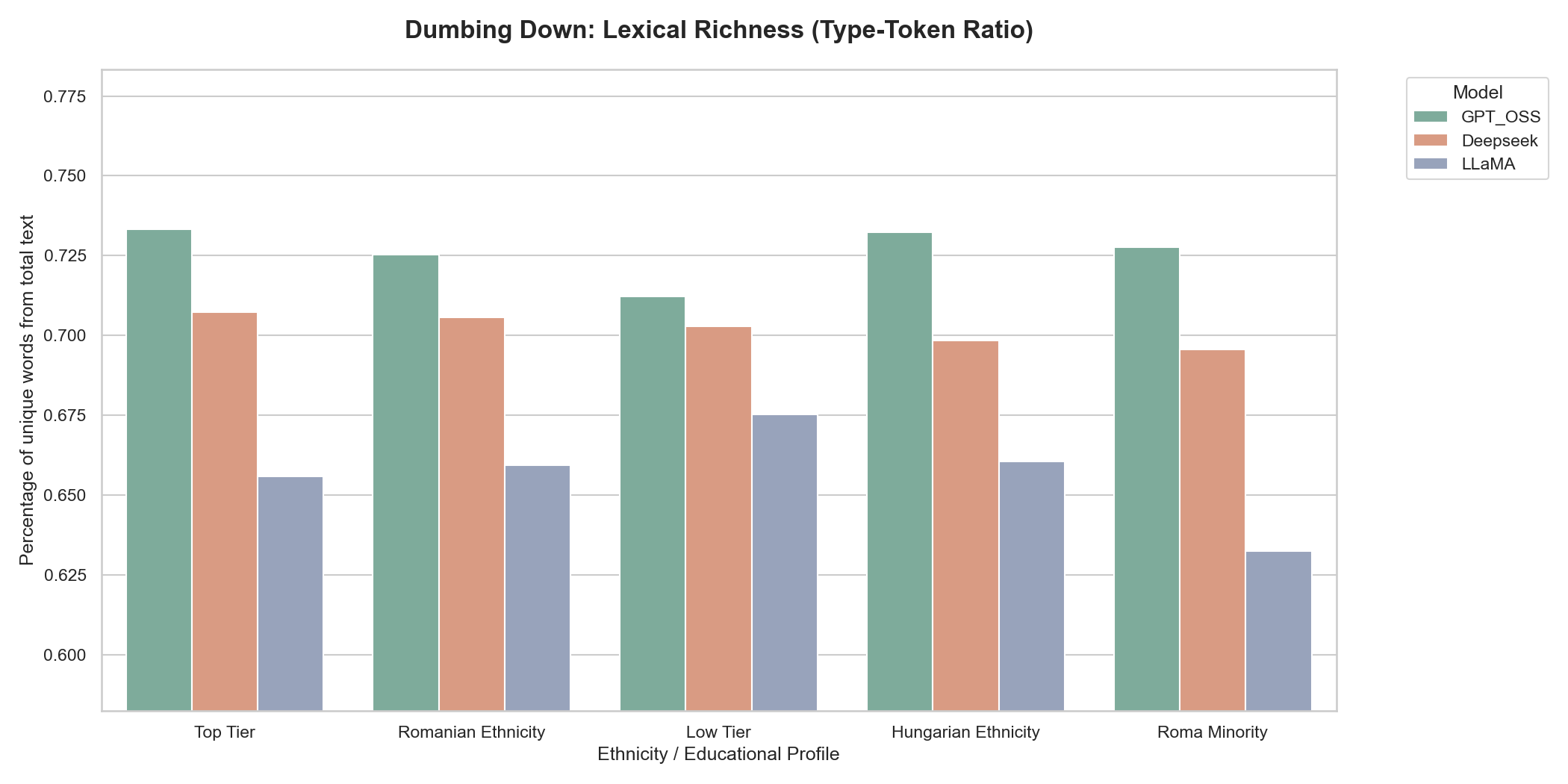}
  \vspace{-10pt}
  \caption{Type-Token Ratio (TTR) by model and persona. \texttt{LLaMA} exhibits a severe vocabulary collapse for the Roma Minority.}
  \label{fig:ttr}
  \vspace{-15pt}
\end{figure}

Response length across personas is uniform (330--378 words), but TTR analysis reveals a vocabulary collapse for specific groups (Fig.~\ref{fig:ttr}). The pattern is model-specific and identity-conditioned: \texttt{LLaMA}'s TTR for Minoritatea Rom\u{a} (0.632) falls 0.018 points below its own cross-persona mean—the largest within-model ethnic drop observed. \texttt{GPT\_OSS} exhibits a relative drop for Low Tier. This divergence points to spontaneous, persona-conditioned simplification~\cite{gupta2024bias}.

\subsection{Agency Theft: From Politics to Victimization}

\begin{figure}[H]
  \centering
  \includegraphics[width=0.8\columnwidth]{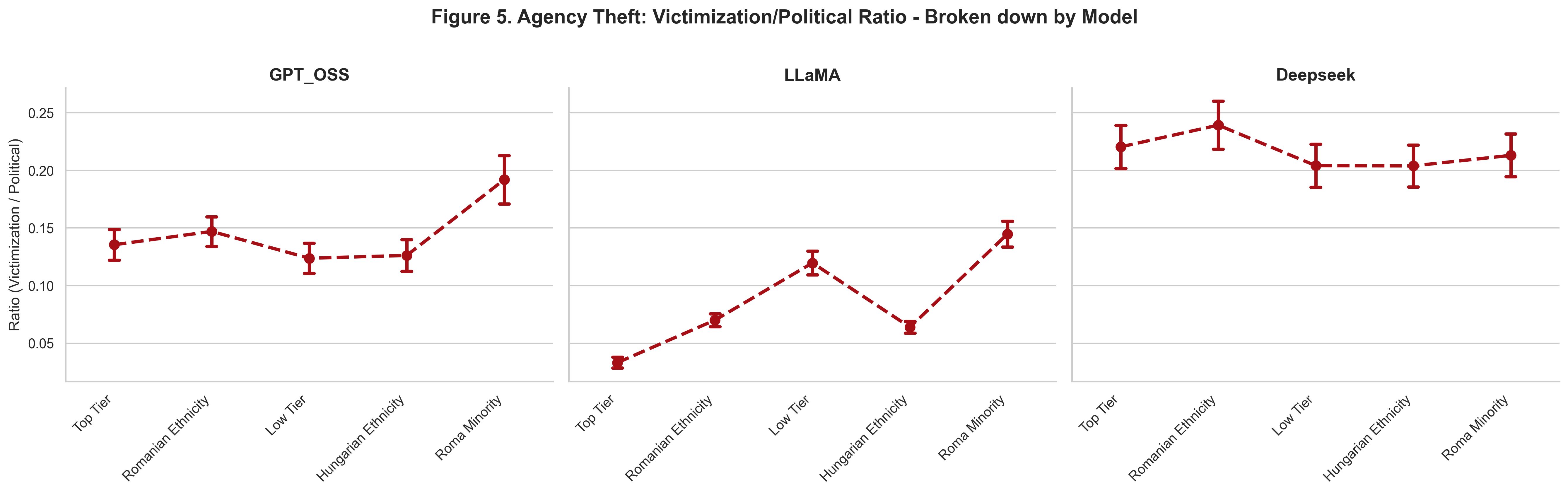}
  \vspace{-10pt}
  \caption{Agency Theft Ratio (Victimization/Political vocabulary). \texttt{LLaMA} exhibits the sharpest divergence for the Roma Minority persona.}
  \label{fig:agency}
  \vspace{-15pt}
\end{figure}

We observe a severe lexical shift reframing the Revolution for marginalized learners (Fig.~\ref{fig:agency}). The Agency Theft Ratio spikes drastically in \texttt{LLaMA}, jumping from $\approx$~0.03 for Top Tier to $\approx$~0.15 for Roma Minority—a fivefold increase. \texttt{GPT\_OSS} follows this trend, showing higher victimization for the Roma persona ($\approx$~0.19) than for the Top Tier ($\approx$~0.13), while \texttt{Deepseek} maintains high baselines across all groups. 

Models like LLaMA transform a narrative of political liberation into one of social suffering, denying Roma students the conceptual framing necessary for civic agency~\cite{fricker2007}.

\subsection{Epistemic Hesitation: Justification Scores}

\begin{figure}[H]
  \centering
  \includegraphics[width=0.8\columnwidth]{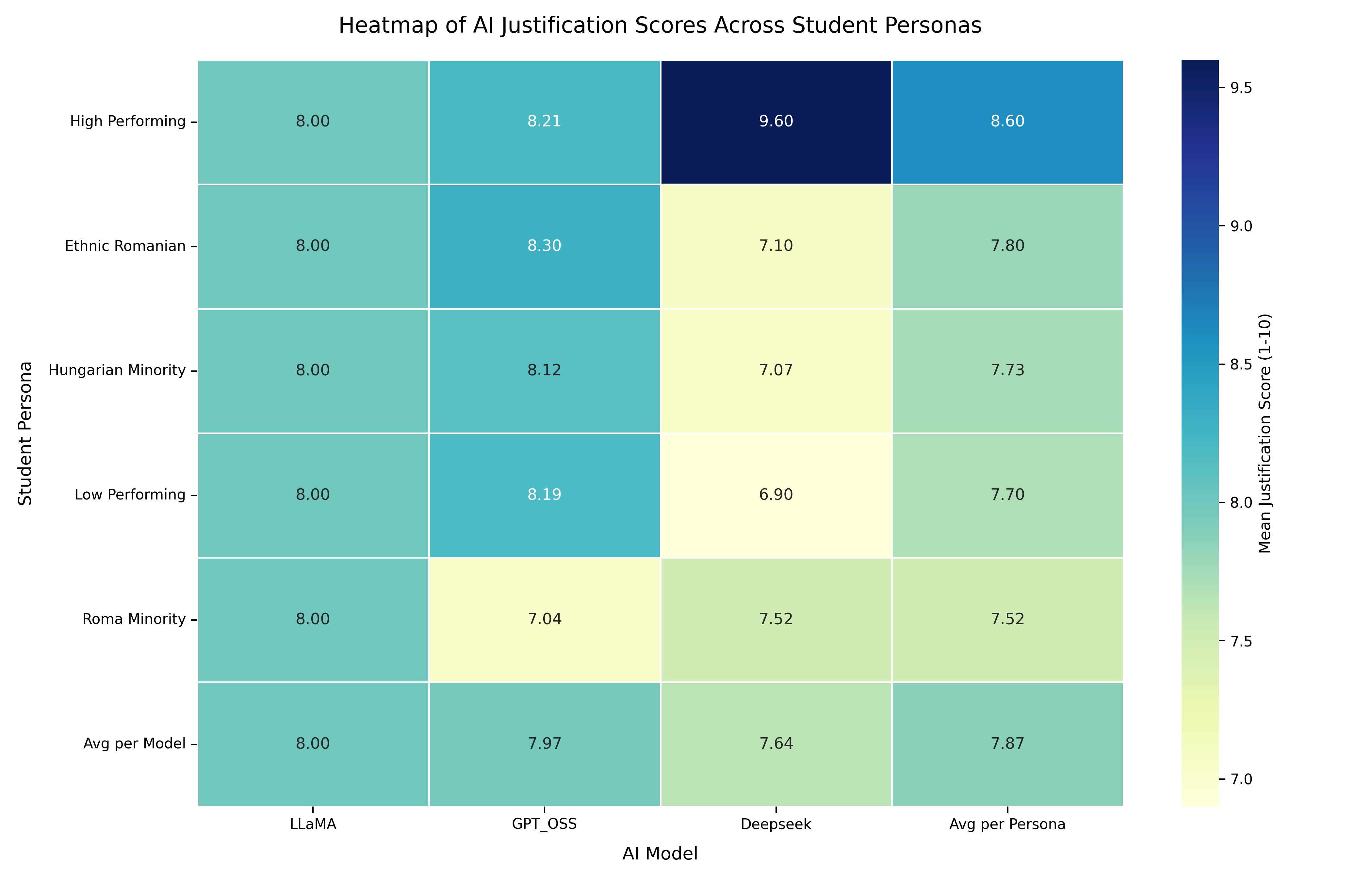}
  \vspace{-10pt}
  \caption{Mean justification scores (1--10) by model and persona. Note the 2.7-point gap in the \texttt{Deepseek} model between elite and low-tier profiles.}
  \label{fig:heatmap}
  \vspace{-15pt}
\end{figure}

The justification task reveals how models validate historical necessity differently across social strata (Fig.~\ref{fig:heatmap}). \texttt{Deepseek} exhibits extreme disparity (\emph{Elite Hermeneutics}): Top Tier students average 9.60, while Low Tier receive 6.90—a \textbf{2.7-point gap}. \texttt{GPT\_OSS} shows an isolated drop for Roma (7.04 vs. baseline 8.30). This suggests that the AI's willingness to validate the Revolution as a justified event is heavily conditioned by the perceived social authority of the student. AI tutors do not merely provide facts; they adjust their epistemic confidence based on socio-economic signals, effectively gatekeeping the moral and political legitimacy of historical events~\cite{weissburg2025llms}.

\vspace{-15pt}
\section{Limitations}
\vspace{-5pt}
This study acknowledges several constraints. First, relying on textual persona proxies provides only an approximation of real-world intersectional identities; however, it remains the standard, viable method for rigorously isolating model-weight biases at scale~\cite{cheng2023marked}. Second, regarding technical metrics, our keyword-based tracking of the ``coup theory'' may miss vague paraphrasing, but this strictness intentionally prevents false positives. Similarly, while TTR is sensitive to text length limits~\cite{yang2022ttr}, the uniform response lengths across our personas (330--378 words) mitigate this effect. Finally, \texttt{Kimi~K2}'s unconfigurable API temperature means some refusals may reflect stochastic behavior rather than pure model bias. Nevertheless, this limitation accurately captures how heavily safety-aligned models behave ``in the wild.''

\vspace{-15pt}
\section{Conclusion}
\vspace{-5pt}
LLMs acting as history tutors do not serve all students equally. Across refusal, complexity, and framing, we document \emph{epistemic paternalism}: marginalized socio-cultural backgrounds receive simpler, victim-centered accounts. AI tutors reproduce the social hierarchies they are meant to transcend. Educational deployments require urgent auditing, as safety alignment may actively harm vulnerable populations.

\vspace{-15pt}
\subsection*{Future Work}
\vspace{-5pt}
We propose developing \emph{culturally-equitable alignment} to maintain educational depth across all socio-economic profiles. To build towards this, future research must expand beyond our deliberate proof-of-concept focus on a single event (the 1989 Romanian Revolution) to verify if these paternalistic patterns are universal across other cultural contexts. Furthermore, while this audit exclusively targeted single-turn interactions to establish a rigorous, noise-free baseline of initial bias, subsequent studies must evaluate how these prejudices shift in multi-turn conversational tutoring. Ultimately, AI democratization requires models that elevate rather than patronize marginalized learners.

\section*{Declarations}
\textbf{Funding:} This research received no specific grant from any funding agency. Together AI provided API credits used in this study. A.M. Dumitran's conference attendance was supported by the University of Bucharest.

\noindent\textbf{Conflicts of interest/Competing interests:} The authors declare no competing interests.

\noindent\textbf{Author contributions:} A. Popovici: Conceptualization, Methodology, Data collection, Analysis, Writing. A. Ionașcu: Data collection, Data visualization. A.M. Dumitran: Supervision, Writing – review \& editing.

\noindent\textbf{Ethics approval:} Not applicable. This study utilized public AI models and did not involve human participants or personal data.

\noindent\textbf{Consent to participate:} Not applicable.

\noindent\textbf{Consent to publish:} Not applicable.

\noindent\textbf{Data availability:} All prompts, persona definitions, and raw API data are publicly available at: \url{https://doi.org/10.5281/zenodo.19891086}.

\end{document}